\documentclass[12pt]{article}

\usepackage[utf8]{inputenc}
\usepackage{geometry}
\usepackage{graphicx}
\usepackage[outdir=./]{epstopdf}
\usepackage{array}
\usepackage{subfig}
\usepackage{slashed}
\usepackage{amsmath}
\usepackage{amsfonts}
\usepackage{stmaryrd}
\SetSymbolFont{stmry}{bold}{U}{stmry}{m}{n} 
\usepackage{amssymb}
\usepackage[cm]{fullpage}
\usepackage{cite}
\usepackage{epsfig}
\usepackage{comment}
\usepackage{hyperref}
\usepackage{multirow}
\usepackage{cancel}
\usepackage{mathrsfs}

\usepackage{cleveref}
\crefname{section}{Â§}{Â§Â§}
\Crefname{section}{Â§}{Â§Â§}

\numberwithin{equation}{section}

\def\p{\partial}

\def\0{{(0)}}
\def\1{{(1)}}
\def\2{{(2)}}

\def\<{\langle }
\def\>{\rangle }

\newcommand{\bea}{\begin{eqnarray}}

\newcommand{\eea}{\end{eqnarray}}

\newcommand{\be}{\begin{equation}}
\newcommand{\ee}{\end{equation}}
\newcommand{\ba}{\begin{align}}
\newcommand{\ea}{\end{align}}

\makeatletter
  \let\over=\@@over \let\overwithdelims=\@@overwithdelims
  \let\atop=\@@atop \let\atopwithdelims=\@@atopwithdelims
  \let\above=\@@above \let\abovewithdelims=\@@abovewithdelims
\renewcommand\section{\@startsection {section}{1}{\z@}%
                                   {-3.5ex \@plus -1ex \@minus -.2ex}
                                   {2.3ex \@plus.2ex}%
                                   {\normalfont\large\bfseries}}

\renewcommand\subsection{\@startsection{subsection}{2}{\z@}%
                                     {-3.25ex\@plus -1ex \@minus -.2ex}%
                                     {1.5ex \@plus .2ex}%
                                     {\normalfont\bfseries}}
\makeatother

\DeclareMathOperator{\SU}{SU}

\newcommand{\beq}{\begin{equation}}
\newcommand{\eeq}{\end{equation}}
\newcommand{\beqa}{\begin{eqnarray}}
\newcommand{\eeqa}{\end{eqnarray}}
\newcommand{\beqar}{\begin{eqnarray*}}

\newcommand{\mc}[1]{\mathcal{#1}}

\def\[{\[}
\def\]{\]}

\newcommand{\bd}[1]{\begin{fmffile}{#1}\begin{fmfgraph*}}
\newcommand{\ed}{\end{fmfgraph*}\end{fmffile}}

\newcommand{\mN}{\mathcal{N}}
\newcommand{\N}{\mathcal{N}}
\newcommand{\mb}{\mathbb}

\newcommand{\Nstar}{\mN=2^\star}

\usepackage{slantsc}
\usepackage{booktabs}

\begin{document}

\pagenumbering{Alph} 
\begin{titlepage}

\unitlength = 1mm~
\vskip 2cm
\begin{center}

{\textbf{\LARGE{\textsc{The Decoupling of {\sc $\bar\Omega$} in String Theory}}}}

\vspace{0.8cm}
\Large{K.S.~Narain\,\footnote{{\tt narain@ictp.it}}~ and A. Zein Assi\,\footnote{\tt zeinassi@cern.ch}}

\begin{center}
\normalsize{
\renewcommand{\thefootnote}{\fnsymbol{footnote}}\vspace{0.5cm}
${}^{1}$ High Energy Section, The Abdus Salam International Center for Theoretical Physics,
\\Strada Costiera, 11-34014 Trieste, Italy\\[0.2cm]
${}^{2}$ Albert Einstein Center for Fundamental Physics\\Institute for Theoretical Physics, University of Bern\\Sidlerstrasse 5, CH-3012 Bern, Switzerland}\\[0.2cm]
\end{center}

\begin{abstract}
In this note, we study the deformation of the topological string by $\bar\Omega$. Namely, adopting the perturbative string amplitudes approach, we identify the $\bar\Omega$-deformation in terms of a physical state in the sting spectrum. We calculate the topological amplitudes $F_g$ in heterotic string theory in the presence of the latter. In particular, we show that it is crucial to include quadratic terms in the effective action in order for $\bar\Omega$ to decouple. It turns out that this decoupling happens at the full string level, suggesting that this holds non-perturbatively.

\end{abstract}

\vspace{1.0cm}

\end{center}

\end{titlepage}

\pagenumbering{arabic} 

\pagestyle{empty}
\pagestyle{plain}

\def\vx{{\vec x}}
\def\p{\partial}
\def\po{$\cal P_O$}

\pagenumbering{arabic}

\tableofcontents

\section{Introduction}

The interplay between string theory and supersymmetric gauge theories has been very fruitful in the past decades in unravelling new connections and structures both in string theory and gauge theory. The typical example is the famous TST-SGT-ST golden triangle. Namely, topological string theory (TST) can be thought of as a sub-sector of string theory (ST) capturing only its BPS states and, on the other hand, is a useful framework to engineer supersymmetric gauge theories (SGT). For instance, the partition function of the $\Omega$-deformed gauge theory \cite{Losev:1997wp,Nekrasov:2002qd,Nekrasov:2003rj} corresponds, in the topological limit, to the topological string theory partition function which calculates a class of gravitational couplings $F_g$ in the string effective action \cite{Antoniadis:1993ze}.

From the gauge theory point of view, in the topological limit, the $\Omega$-deformation consists of a one-parameter twist of space-time denoted $\epsilon$ which lifts up to the topological string coupling $g_s^{\textrm{top}}$. From the effective action point of view, the latter is a constant background for the self-dual graviphoton field strength.

The topological nature of $F_g$ implies that this coupling depends holomorphically on some of the moduli of the Calabi-Yau compactifications, modulo anomalous, boundary terms \cite{Bershadsky:1993cx} that are absent in the low energy gauge theory\footnote{For a calculation of these non-holomorphic terms in supergravity see \cite{Cardoso:2010gc}.}. In the latter, additionally, since the $\Omega$-deformation can be thought of as a geometrical twist of space-time \cite{Nekrasov:2003rj}, one can promote $\epsilon$ to a complex parameter and supersymmetry implies that the gauge theory partition function is holomorphic in $\epsilon$, \emph{i.e.} independent of $\bar\epsilon$.

Our goal is to study whether there is a notion of holomorphy in $\epsilon$ in string theory. From the TST side, this is not natural since $\epsilon$ is $g_s^{\textrm{top}}$. Therefore, we analyse the question from the effective action point of view. Namely, we first identify an extension of $F_g$ that captures also $\bar\epsilon$. More precisely, we give an exact identification of the latter in terms of physical fields\footnote{Similar questions were analysed in \cite{Billo:2006jm} from a low energy perspective.}. The complete proof of this identification is the subject of \cite{OvsG}.

We first start by identifying $\bar\epsilon$ in Section \ref{DefOmegaBar} in heterotic string theory through an intuitive argument based on the geometrical picture of the underlying gauge theory. Then, in Section \ref{Het}, we perform an explicit calculation of the coupling $F_g$ including the $\bar\Omega$-deformation and analyse the decoupling of the latter. In particular, we show that it is important to include quadratic terms in the deformation in order not only to ensure holomorphy in $\epsilon$ but also to obtain the correct gauge theory partition function. Useful results and technical details are deferred to two appendices.


\section{\texorpdfstring{What is $\bar\Omega$}{What is OmegaBar}}\label{DefOmegaBar}

In order to identify the $\bar\Omega$-deformation in terms of the physical string spectrum, we consider for simplicity the heterotic $E_8\times E_8$ string theory compactified on $T^2\times K3$ where $K3$ is realised as a $\mb Z_2$ orbifold of $T^4$. In fact, our results do not depend on the particular realisation of $K3$ which could even be taken to be non-compact. We focus on the vector multiplet sector which consists in the so-called STU-model. Namely, there are three vector multiplet moduli called S, T and U where S is the dilaton while T and U are the K\"ahler and complex structure moduli of the space-time $T^2$. To keep the discussion clear, we do not turn on any Wilson lines. In addition to the three vector fields associated to each scalar and which we denote as $F_I$ with $I=S,T,U$, there is another vector field stemming from the $\N=2$ gravity multiplet, \emph{i.e.} the graviphoton, which we denote as $F_G$ subsequently.

Since we are concerned by the realisation of the $\Omega$-deformation, we recall some of its basics facts. First of all, one can realise it geometrically by a non-trivial $T^2$ fibration over space-time considered as $\mb R^4$, such that the space-time fields are rotated with an arbitrary angle whenever one goes around the cycles of $T^2$. Therefore, the $SO(4)$ Lorentz group, whose covering group is $\SU(2)_L\times SU(2)_R$, is explicitly broken. This construction can be thought of as a reduction from six dimensions on the metric with line element\footnote{See \cite{Hellerman:2011mv} for the M-theory uplift of this picture.}
\begin{equation}\label{OmegaMetric}
\textrm{d}s^2=g_{\mu\nu}\left(\textrm{d}Z^\mu+{\Omega^\mu}_\rho Z^\rho \textrm{d}X+{\bar\Omega^\mu}_\rho Z^\rho \textrm{d}\bar X\right)\left(\textrm{d}Z^\nu+{\Omega^\nu}_\rho Z^\rho \textrm{d}X+{\bar\Omega^\nu}_\rho Z^\rho \textrm{d}\bar X\right)+\textrm{d}s^2_{T^2}.
\end{equation}
Here, $\mu,\nu\in[\![0,3]\!]$ are space-time indices, $X$ is the complexified $T^2$ coordinate and $\Omega,\bar\Omega$ are $4\times4$ matrices parametrising the rotations in space-time. In general, they depend on two complex parameters $\epsilon_{1,2}$ which can be redefined as
\begin{equation}
 \epsilon_{\pm}=\frac{\epsilon_1\pm\epsilon_2}{2}
\end{equation}
in order to make manifest to decomposition of the Lorentz group. This background preserves supersymmetry if one includes an R-symmetry rotation of $\epsilon_+$. For the present matter, we work in the topological limit $\epsilon_+=0$. In \cite{Antoniadis:1993ze}, it was shown that the holomorphic part of the $\Omega$-deformation in this limit, \textit{i.e.} $\epsilon_-\equiv\epsilon$, can be described as a constant background for the self-dual part of the graviphoton field strength. The self-duality condition is consistent with the fact that $\epsilon$ is sensitive to $SU(2)_L$ only. In this convention, $SU(2)_R$ remains unbroken. Hence, in order to describe $\bar\Omega$ in string theory, one is led to turn on a constant background for a self-dual field strength other then the graviphoton one. There are thus three choices.

\begin{enumerate}
 \item ${F_T}^{(-)}$: $V_T=\epsilon_\mu\partial Z^\mu\bar\partial\bar X$\,.
 \item ${F_U}^{(-)}$: $V_U=\epsilon_\mu\partial Z^\mu\bar\partial X$\,.
 \item ${F_S}^{(-)}$: $V_S=\epsilon_\mu\partial\bar X\bar\partial Z^\mu$\,.
\end{enumerate}

The superscript $(-)$ refers to the fact that the field strengths are chosen to be self-dual. We have also written the (bosonic part of the) vertex operator of each field, with the understanding that the polarisation $\epsilon_\mu$ is chosen in such a way to satisfy the self-duality constraint. Recall that the graviphoton vertex operator is
\begin{equation}
 V_G=\epsilon_\mu\partial X\bar\partial Z^\mu\,.
\end{equation}
Consequently, the two natural choices for $\bar\Omega$ are ${F_T}^{(-)}$ and ${F_S}^{(-)}$. By `natural' we mean that $\bar\Omega$ should be understood, in some sense, as the complex conjugate of $\Omega$. It turns out that the choice ${F_S}^{(-)}$ is the \emph{wrong} one. Therefore, we claim that \textit{the $\bar\Omega$-deformation is realised as a constant background for $F_T^{-}$}.

This is shown fully non-perturbatively in \cite{OvsG} by an explicit derivation of the $\bar\Omega$-deformed gauge theory and ADHM effective actions. In what follows, we provide further support by showing not only that the decoupling happens in string theory but also that it is \emph{exact}.


\section{\texorpdfstring{$\bar\Omega$ decoupling from topological amplitudes}{OmegaBar decoupling from topological amplitudes}}\label{Het}

\subsection{Effective action}

Consider the following series of effective couplings in the standard four-dimensional superspace $\mathbb{R}^{4|8}\sim\{x^\mu,\theta^i_\alpha,\bar{\theta}_i^{\dot{\alpha}}\}$ \cite{Antoniadis:1993ze}:
\begin{align}
\mathcal{I}_g=&\int d^4x \int d^4\theta\, \mathcal{F}_g(X)\,(W_{\mu\nu}^{ij} W_{ij}^{\mu\nu})^g &&\text{for} && g\geq1\,,  \label{BpsStandard}
\end{align}
which is a $\frac{1}{2}$-BPS F-term since it is invariant under half of the supercharges. In addition, $W_{\mu\nu}^{ij}$ is the supergravity multiplet and we have introduced $SU(2)_R$ R-symmetry group indices $i,j=1,2$. The superfield $W_{\mu\nu}$ contains the graviphoton field-strength $F^G$, the field strength tensor $B_{\mu\nu}^i$ of a doublet of gravitini and the Riemann tensor:
\begin{align}
W_{\mu\nu}^{ij}=F^{G
,ij}_{(-),\mu\nu}+\theta^{[i} B_{(-),\mu\nu}^{j]}-(\theta^i\sigma^{\rho\tau}\theta^j) R_{(-),\mu\nu\rho\tau} + \cdots\,
\end{align}
The subscript $(-)$ denotes the self-dual part of the corresponding field strength tensor. The coupling function $\mathcal{F}_g$ in (\ref{BpsStandard}) only depends on holomorphic vector multiplets, which contain a complex scalar $\phi$, an $SU(2)_R$ doublet of chiral spinors $\lambda_\alpha^i$ as well as a self-dual field-strength tensor of a space-time vector $F^{\mu\nu}_{(-)}$:
\begin{align}
X^I=\phi^I+\theta^i\lambda_i^I+\tfrac{1}{2}F_{(-)\,\mu\nu}^I \epsilon_{ij}(\theta^i\sigma^{\mu\nu}\theta^j) +\cdots\,
\end{align}
The index $I$ labels the different vector multiplets. One of them, denoted $X^0$, is unphysical and serves as a compensator of degrees of freedom in the formulation of ${\mc N}=2$ supergravity. The physical moduli are then the lowest components of the projective multiplets:
\begin{align}
\hat{X}^I:= \frac{X^I}{X^0}\,.
\end{align}
Upon explicitly performing the integral over the Grassmann variables, (\ref{BpsStandard}) induces a component term
\begin{align}\label{BpsTerm}
\mathcal{I}_g=\int d^4x\, \mathcal{F}_g(\phi)\,R_{(-)\,\mu\nu\rho\tau}R_{(-)}^{\mu\nu\rho\tau}\,
\left[F^G_{(-)\,\lambda\sigma}F^{G\,\lambda\sigma}_{(-)}\right]^{g-1}+\cdots\,
\end{align}

We now modify this coupling in order to allow for an insertion of an arbitrary number of additional self-dual field strength. To be precise, consider the descendent multiplet
\begin{equation}
 K_{\mu\nu}=\left(\epsilon_{ab}D^a\sigma_{\mu\nu}D^b\right)X=F_{(-)\mu\nu}+\cdots\,,
\end{equation}
where the subscript $(-)$ recall that we are concerned by self-dual fields only, and insert it in the effective action term \eqref{BpsStandard} as
\begin{align}
\mathcal{I}_{g,n}=&\int d^4x \int d^4\theta\, \mathcal{F}_{g,n}(X)\,W^{2g} K^{2n} \,.  \label{BpsNew}
\end{align}
In the subsequent sections, we calculate the coupling function $F_{g,n}$ for arbitrary $g$ and $n$ in heterotic string theory compactified on $T^2\times K3$. As explained above, we choose $K$ to be the descendent multiplet of the K\"ahler structure $T$ of $T^2$ in heterotic string theory.

\subsection{Amplitude calculation}

In order to calculate the coupling $F_{g,n}$, recall that the standard topological couplings $F_g$, in heterotic string theory, receive perturbative contributions at one-loop only due to the fact that the heterotic dilaton is a vector multiplet scalar satisfying a Peccei-Quinn symmetry. Therefore, the new couplings \eqref{BpsNew} are also naturally calculated perturbatively at one-loop.

For convenience, we choose a particular kinematic configuration in which the states of interest (gravitons, graviphotons and $T$-vectors) carry space-time momenta along $Z^1$ and $\bar Z^2$ only. The vertex operators for the gravitons are
\begin{align}
 V_{R}(x)&=(\partial Z^2-i\bar p_1\psi^1\psi^2)\bar\partial Z^2\,e^{i\bar p_1 Z^1}\,,\nonumber\\
 V_{R}(y)&=(\partial \bar Z^1-i p_2\bar\psi^2\bar\psi^1)\bar\partial\bar Z^1\,e^{i p_2\bar Z^2}\,,
\end{align}
while for the graviphotons these are
\begin{align}
 V_{G}(z_i)&=(\partial X-i\bar p_1\psi^2\chi)\bar\partial Z^2\,e^{i\bar p_1 Z^1}\,,\nonumber\\
 V_{G}(w_i)&=(\partial X-ip_2\bar\psi^2\chi)\bar\partial\bar Z^1\,e^{ip_2\bar Z^2}\,.
\end{align}
Finally, for the $T$-vectors, the vertex operators are
\begin{align}
 V_{T}(s_i)&=(\partial Z^2-i\bar p_1\psi^1\psi^2)\bar\partial\bar X\,e^{i\bar p_1 Z^1}\,,\nonumber\\
 V_{T}(t_i)&=(\partial \bar Z^1-i p_2\bar\psi^2\bar\psi^1)\bar\partial\bar X\,e^{i p_2\bar Z^2}\,.
\end{align}

Consider the amplitude
\begin{equation}\label{AmpBarBis}
 \left\langle V_R(x)V_R(y)\prod_{i=1}^{g-1} V_G(z_i)V_G(w_i)\prod_{j=1}^n V_{T}(s_i)V_{T}(t_i)\right\rangle\,.
\end{equation}
It turns out that it is technically simpler to consider the supersymmetrically related amplitude where, instead of the component \eqref{BpsTerm}, we take in \eqref{BpsNew} the graviphoton field strength from all the supergravity multiplet and extract two $F_T$ field strengths from the coupling function. These are the higher components of the $T$-modulus chiral multiplet and they come with an normalisation of $1/T_2$. Therefore, in the following, it is understood that one has to include a factor of $1/(T_2)^2$. On the other hand, the vertices stemming from $W$ and $K$ must be multiplied by $e^{K_0/2}$ where $K_0$ is the K\"ahler potential stripped off its dilaton term. Hence, we focus on the amplitude
\begin{equation}\label{AmpBar}
 \left\langle V_T(x)V_T(y)\prod_{i=1}^{g} V_G(z_i)V_G(w_i)\prod_{j=1}^n V_{T}(s_i)V_{T}(t_i)\right\rangle\,.
\end{equation}
This amplitude calculates the second derivative of the coupling $F_{g,n}$ with respect to the modulus $T$.

With this remark in mind, we use the standard generating function trick to calculate all amplitudes of interest as
\begin{equation}\label{GenFunction}
 \partial_T^2F(\epsilon,\bar\epsilon)=\sum_{g,n}\frac{\epsilon^{2g}\bar\epsilon^{2n}}{g!^2 n!^2}\left\langle V_T(x)V_T(y)\prod_{i=1}^{g} V_G(z_i)V_G(w_i)\prod_{j=1}^n V_{T}(s_i)V_{T}(t_i)\right\rangle\,.
\end{equation}
In the previous equation, we have chosen that the space-time zero-modes are soaked up by the vertices at $x$ and $y$. This is of course arbitrary and one could have taken any other two vertices $V_T$ instead, which leads to the same term upon relabeling the positions. This leads to a combinatorial factor of $(n+1)^2$ already taken into account in \eqref{GenFunction}. Furthermore, the integration over the positions is implicitly understood and we factor out a momentum power $(\bar p_1\,p_2)^{g+n+1}$ from every term in \eqref{GenFunction}.
We first focus on the fermionic contractions in a particular term where a number $2N$ $F_T$ vertices give their fermionic parts. In order to perform the spin-structure sum and soak up the $T^2$ zero-modes in the odd spin structure, we take one of the graviphotons to be in the (-1)-picture and insert one PCO. The latter necessarily gives its $T_F$ part in the $T^2$ directions. Since no other terms can give fermionic terms in $T^2$, the superghost partition function cancels that of the $T^2$ and we are left with
\begin{align}
 \theta_s\left(x-y+\sum_{j}(s_i-t_i)\right)&\theta_s\left(x-y+\sum_j(s_j-t_j)\right)\theta_{s,h}(0)\theta_{s,-h}(0)\nonumber\\
&\times\frac{\prod_{j=1}^N\theta_1(x-s_j)^2\theta_1(y-t_j)^2\prod_{i<j=1}^N\theta_1(s_i-s_j)^2\theta_1(t_i-t_j)^2}{\theta_1(x-y)^2\prod_{j=1}^{N}\theta_1(x-t_j)^2\theta_1(y-s_j)^2}\,,
\end{align}
where $s$ is the spin structure and $h$ denotes the twist along $K3$ which we take to lie at an orbifold point for simplicity. The spin-structure sum can readily be made, leading to
\begin{align}\label{sssum}
 \theta_1\left(x-y+\sum_{j}(s_i-t_i)\right)&\theta_1\left(x-y+\sum_j(s_j-t_j)\right)\theta_{h}(0)\theta_{-h}(0)\nonumber\\
&\times\frac{\prod_{j=1}^N\theta_1(x-s_j)^2\theta_1(y-t_j)^2\prod_{i<j=1}^N\theta_1(s_i-s_j)^2\theta_1(t_i-t_j)^2}{\theta_1(x-y)^2\prod_{j=1}^{N}\theta_1(x-t_j)^2\theta_1(y-s_j)^2}\,.
\end{align}
Notice that only the twisted sectors of $K3$ are non-zero. Notice that \eqref{sssum} is nothing but the correlation function, in the odd spin structure, of $(N+1)$ current $\psi^1\psi^2$ at positions $x$ and $s_i$ with $(N+1)$ currents $\bar\psi^1 \bar\psi^2$ at positions $y$ and $t_i$. Hence, upon soaking up the space-time zero-modes, we can write \eqref{sssum} as a correlator in the odd spin structure with the zero-modes stripped off:
\begin{equation}
 \left\langle\prod_{j=1}^{N}\psi^1\psi^2(s_j)\bar\psi^1\bar\psi^2(t_j)\right\rangle'\,.
\end{equation}

We now turn to the bosonic terms and consider the case where $2M$ operators $F_T$ give their bosonic terms only. Notice that the bosonic operators naturally bring down a momentum power from the exponential of each vertex operator in such a way that the amplitude calculates \eqref{BpsNew}. Factoring out the space-time momentum power, we obtain the correlation function
\begin{align}
 \left\langle Z^1\partial Z^2\bar\partial\bar X(x)\bar Z^2\partial\bar Z^1\bar\partial\bar X(y)\prod_{i=1}^{g}\partial XZ^1\bar\partial Z^2(z_i)\partial X\bar Z^2\bar\partial\bar Z^1(w_i)\prod_{j=1}^{M}Z^1\partial Z^2\bar\partial\bar X(s_j)\bar Z^2\partial\bar Z^1\bar\partial\bar X(t_j)\right\rangle\,.
\end{align}
We now strip off a factor of $(\partial X\bar\partial\bar X)^2$ which we can simply interpret as a marginal deformation that realises the second derivative with respect to $T$. Consequently, including all possible partitions of $n=M+N$ and summing over $n$ and $g$, we find that the generating function \eqref{GenFunction} can be written in the elegant form
\begin{align}\label{GenSim}
 F(\epsilon,\bar\epsilon)=\Bigg\langle \textrm{exp}\Big[&-\epsilon\int\textrm{d}^2z\,\left(\partial XZ^1\bar\partial Z^2+\partial X\bar Z^2\bar\partial\bar Z^1\right)\nonumber\\
&-\bar\epsilon\int\textrm{d}^2z\,\left((Z^1\partial Z^2+\psi^1\psi^2)\bar\partial\bar X+(\bar Z^2\partial\bar Z^1+\bar\psi^1\bar\psi^2)\bar\partial\bar X\right)\Big]\Bigg\rangle\,.
\end{align}
This can be viewed as a deformed worldsheet sigma-model of which we are calculating the one-loop partition function in the odd spin structure and with the fermionic zero-modes already soaked up. It is important to mention that, on top of their zero-modes, the bosonic fields in $T^2$ can contract among themselves giving rise to terms which are irrelevant for the field theory limit. However, since we are interested in the question of decoupling of $\bar\Omega$ at the string level, we keep track of these contact terms.

\subsection{Inclusion of quadratic terms}

In order to calculate \eqref{GenSim} exactly, we first show that it is indeed Gaussian. As mentioned previously, in the correlation function leading to $F_{g,n}$, the compact bosons $\partial X$ and $\bar\partial\bar X$ can either give their zero-modes denoted by $Q_L$ and $\bar Q_L$ respectively or contract as dictated by the torus propagator. $Q_{L,R}$ are the $T^2$ lattice momenta in the Lagrangian representation which is the natural representation arising from the path integral, see Appendix \ref{appendix:lattice}. In order to go to the more physical Hamiltonian representation, we perform a double Poisson re-summation. This is explained in detail in Appendix \ref{appendix:lattice} where all the relevant definitions and technical details are summarised. The important result for us is that, upon going to the Hamiltonian representation, the correlation function $\left<\partial X\bar\partial\bar X\right>$ becomes $P_L\bar P_R$, namely
\begin{align}
 \frac{T_2}{\tau_2}\sum_{\tilde m^i,n^i}\left(\left<\partial X\bar\partial\bar X\right>\right)^k q_T^{|Q_L|^2}\bar q_{\bar T}^{|Q_R|^2}=\sum_{m_i,n^i}\left(-P_L\bar P_R\right)^k q^{|P_L|^2}\bar q^{|P_R|^2}\,.
\end{align}
This means that, once we go to the Hamiltonian representation, the generating function \eqref{GenSim} becomes effectively Gaussian and one can evaluate it exactly. This gives rise to an infinite product which can be regularised using analytic continuation through Eisenstein series. The explicit evaluation is rather technical and is presented in Appendix \ref{appendix:reg}. For our present matter, we focus on the modes surviving the field theory limit. These are the $n=0$ modes of the Laplacian eigenvalues expansion. The result is
\begin{align}\label{wrongres}
 F(\epsilon,\bar\epsilon)&=\int\frac{\textrm{d}t}{t}\left(\frac{\sin(\pi\hat{\bar\epsilon} t)}{\pi\bar\epsilon}\right)^2\left(\frac{\pi(\epsilon+\bar\epsilon)}{\sin(\pi(\tilde\epsilon+\hat{\bar\epsilon}))}\right)^2 \textrm{e}^{-|P_L|^2 t}\,.
\end{align}
Here $\tilde\epsilon$ and $\hat{\bar\epsilon}$ are the deformation parameters dressed with a single power of $T^2$ zero-modes $P_L$ and $\bar P_R$ respectively with $\tau_2$ replaced by $t$, and these originate from the bosonic current of $T^2$:
\begin{align}
 \tilde\epsilon&\equiv\frac{\tau_2 P_L}{(T-\bar T)(U-\bar U)}\epsilon\,,\\
 \hat{\bar\epsilon}&\equiv\frac{\tau_2 \bar P_R}{(T-\bar T)(U-\bar U)}\bar\epsilon\,.
\end{align}
Notice that this is \emph{not} the correct field theory limit and the parameter $\bar\Omega$, naively, does not decouple from the partition function even in the field theory limit! The reason is quite clear. One must take into account the quadratic terms in the deformation parameters, similarly to what happens in field theory \cite{OvsG}. Indeed, these terms are essential in order to decouple the $\bar\Omega$-dependent terms in the field theory effective actions as $Q$-exact terms.

In order to realise this feature, we turn on the quadratic deformation
\begin{equation}\label{quaddef}
 \delta S_2=\epsilon\bar\epsilon\int\textrm{d}^2z\,|Z^i|^2\partial X\bar\partial\bar X
\end{equation}
in the deformed sigma model \eqref{GenSim} and repeat the same calculation as before. Already at the string level, there is a great simplification between bosonic and fermionic degrees of freedom. Namely, due to the quadratic term, there is a complete left-right factorisation in the partition function - see Appendix \ref{appendix:reg} - such that the space-time left-moving bosons (\emph{i.e.} $\tau$-dependent part) and fermions partition functions cancel between themselves. To be precise, the path integral over the bosonic degrees of freedom, including $\delta S_2$ is
\begin{align}
F_{\textrm{bos,mod}}&=\prod_{(m,n)\neq(0,0)}\left(|m+n\tau|^2+(m+n\tau)\tilde\epsilon+(m+n\bar\tau)\hat{\bar\epsilon}+\tilde\epsilon\hat{\bar\epsilon}\right)^2\nonumber\\
                     &=\prod_{(m,n)\neq(0,0)}\left(m+n\bar\tau+\tilde\epsilon\right)^2\left(m+n\tau+\hat{\bar\epsilon}\right)^2\,.
\end{align}
This is to be contrasted with the expression for the path integral without the quadratic deformation given in Appendix \ref{appendix:reg}. As for the fermionic part of the path integral, it is not modified and leads to the mode expansion
\begin{equation}
 F_{\textrm{bos,mod}}=\prod_{(m,n)\neq(0,0)}\left(m+n\tau+\hat{\bar\epsilon}\right)^2\,,
\end{equation}
such that the Bose-Fermi cancellation is obvious even before regularising the infinite products.

The full amplitude can then be written by including the internal and gauge degrees of freedom:
\begin{align}\label{FullAmplitudeHet}
 F\textrm(\epsilon,\bar\epsilon)&=\int_{\mathcal{F}}\frac{d^2\tau}{\tau_2}\,\left(\frac{2\pi\epsilon\bar\eta^3}{\bar\theta_1(\pi\tilde\epsilon)}\right)^2\textrm{e}^{-\frac{\pi}{\tau_2}\tilde\epsilon^2}\frac{1}{\eta^4\bar\eta^{24}}\frac{1}{2}\sum_{h,g=0}^1 Z[^h_g]~\Gamma_{(2,2+8)}(T,U,Y)\,,
\end{align}
where 
\begin{align}\label{Zblock}
	Z[^h_g] = \Gamma_{K3}[^h_g]~\frac{1}{2}\sum\limits_{k,\ell=0,1}\bar\theta^6[^k_\ell]\bar\theta[^{k+h}_{\ell+g}]
\bar\theta[^{k-h}_{\ell-g}]~,
\end{align}
is the K3-lattice together with the partition function of $E_7\times SU(2)$, since one of the $E_8$-group factors is broken by the $\mathbb{Z}_2$-orbifold. The K3-lattice is given explicitly by
\begin{align}\label{K3lattice}
	\Gamma_{K3}[^h_g] ~=~ \Biggr\{ \begin{array}{l l}
							\Gamma_{(4,4)}(G,B) & ,~(h,g)=(0,0) \\
							\left|\frac{2\eta^3}{\theta[^{1+h}_{1+g}]}\right|^4 & ,~(h,g)\neq(0,0) \\
							       \end{array} ~.
\end{align}
For convenience, we have combined the $T^2$- and $E_8$- lattices into $\Gamma_{(2,2+8)}$. Notice that \eqref{FullAmplitudeHet} is nothing but the result of \cite{Antoniadis:1995zn} for the generating function of the $F_g$ amplitudes. In particular, it is independent of $n$. Hence, taking its field theory limit at a gauge symmetry enhancement point, \emph{e.g.} the $T=U$ point, we recover the standard Barnes double gamma function describing the perturbative part of the unrefined Nekrasov partition function:
\begin{equation}
 F^{\textrm{Nek}}_{\textrm{pert}}=\int\frac{\textrm{d}t}{t}\frac{\epsilon^2}{\sin^2(\epsilon t)}e^{-\mu t}\,,
\end{equation}
where $\mu=\sqrt{\frac{i}{\pi}}(T-U)$ is the mass of the BPS state becoming massless at the enhancement point and parametrises the Coulomb branch of the gauge theory.

Before closing this section, we would like to comment on the inclusion of the quadratic deformation \eqref{quaddef}. The latter can be understood as the vertex operator of the $T$-modulus of $T^2$ at second order in the space-time momenta with a vacuum expectation value given by $\epsilon\bar\epsilon$. This means that our starting point is a deformation of the BPS coupling \eqref{BpsNew} with an arbitrary power of the chiral field $X$ corresponding to the $T$-modulus. Hence, in the amplitude \eqref{AmpBar}, we should include an arbitrary number $m$ of the corresponding scalar vertex operator. At fixed power of $\epsilon$ and $\bar\epsilon$, the latter, can also contribute its zero-momentum part. However, this is nothing but a marginal deformation of the amplitude $F_{g-m,n-m}$. Therefore, this contribution must be subtracted and we only include the term with two space-time momenta. Finally, to define a generating function, we sum over the powers of all vertex operators with the appropriate combinatorial factors and this leads to the deformed sigma-model \eqref{GenSim} with the quadratic deformation \eqref{quaddef}. It would be interesting to understand this point from the supergravity point of view.


\section{Conclusions}\label{Conclusion}

In this note, we have realised the $\bar\Omega$-deformation of gauge theory in string theory as a constant background for a physical field. More precisely, focusing on heterotic string theory compactified on $T^2\times K3$, we argued that $\bar\Omega$ can be understood as a constant background for the self-dual vector partner of $T$, the K\"ahler structure of $T^2$. In our convention, $\Omega$ is realised as the self-dual graviphoton field strength.

In order to support our idea, we calculated the topological amplitudes $F_g$ in the presence of the $\bar\Omega$-deformation. In heterotic string theory, this is a one loop amplitude capturing the topological string partition function to all genera in the large base limit of the elliptically-fibered Calabi-Yau manifold of the dual type II string theory \cite{Marino:1998pg}. We show that it is important to take into account a quadratic deformation of the effective action of the form $\Omega\bar\Omega$. The latter can be understood as a background for the $T^2$ metric. This is very similar to what was observed in \cite{Micha} in the context of the realisation of $\Nstar$ in string theory \cite{Florakis:2015ied} and should be expected due to the similar nature of $\Omega$ and the mass deformation. Here, we find that, in the presence of $\bar\Omega$, $F_g$ can still be calculated exactly at one loop and is independent of $\bar\Omega$. In other words, at least perturbatively, $\bar\Omega$ decouples from the topological string partition function.

Since the heterotic result captures parts of the topological string partition function to all genera, it is tempting to conjecture that the decoupling holds beyond the large base limit. It would be interesting to analyse this question directly in the dual type II string theory where the amplitude $F_g$ is perturbatively exact at genus $g$.

Besides, we have presently only focused on the topological limit of the $\Omega$-deformation. A natural subsequent problem is to extend the analysis to the full $\Omega$-deformation using the generalised couplings studied in \cite{AFHNZ, Antoniadis:2013mna}. In that setting, generically, the decoupling is not expected to be obvious due to the contamination of non-BPS states. We plan to come back to these questions in the future.

Finally, the extension of the proof of the $\bar\Omega$ realisation in string theory to the full non-perturbative level is the subject of a future publication \cite{OvsG}.


\section*{Acknowledgements}

A.Z.A would like to thank the ICTP, Trieste for its hospitality during the accomplishment of this project. The work of A.Z.A is supported by the Swiss National Science Foundation.

\appendix

\section{Lattice sums}\label{appendix:lattice}

In this section, we present useful results for the derivation of \eqref{GenSim} on the $T^2$-lattice partition function with K\"ahler modulus $T$ and complex structure $U$. In the Lagrangian representation, it is given by
\begin{equation}\label{LagLat}
 \Gamma_{2,2}=\frac{T_2}{\tau_2}\sum_{\tilde m_i,n_i}q_{T}^{|Q_L|^2}\bar q_{\bar T}^{|Q_R|^2}\,,
\end{equation}
where, $q_T=e^{2i\pi T}$. The lattice momenta are defined as
\begin{align}\label{QL}
 Q_L&=\frac{1}{2i\sqrt{\tau_2 U_2}}\left(\tilde m_1+\bar U\tilde m_2-\bar\tau(n_1+\bar U n_2)\right)\,,\\
 Q_R&=\frac{1}{2i\sqrt{\tau_2 U_2}}\left(\tilde m_1+\bar U \tilde m_2-\tau(n_1+\bar U n_2)\right)\,.\label{QR}
\end{align}
Using the relation $|Q_L|^2-|Q_R|^2=\tilde m_1n_2-\tilde m_2n_1$, \eqref{LagLat} can be written as
\begin{align}
 \Gamma_{2,2}=\frac{T_2}{\tau_2}\sum_{\tilde m_i,n_i}e^{2i\pi\bar T(\tilde m_1n_2-\tilde m_2n_1)-4\pi T_2|Q_L|^2}\,.
\end{align}
In order to go to the Hamiltonian representation, we perform a double Poisson re-summation on the winding modes $\tilde m_i$, \emph{i.e} a discrete Fourier transformation in these modes given by the general formula
\begin{equation}\label{poisson}
 \sum_{m^{i}\in\mb{Z}}e^{-\pi A_{ij}m^im^j+\pi B_im^i}=\frac{1}{\sqrt{\textrm{det}A}}\sum_{\tilde m_i\in\mb{Z}}e^{-\pi(\tilde m_i+\frac{i}{2}B_i)(A^{-1})^{ij}(\tilde m_j+\frac{i}{2}B_j)}\,.
\end{equation}
This leads to the Hamiltonian representation of the lattice
\begin{align}\label{HamLat}
 \Gamma_{2,2}=\sum_{m^i,n_i}q^{|P_L|^2}\bar q^{|P_R|^2}=\sum_{m^i,n_i}\bar q^{n\cdot m}e^{-4\pi\tau_2|P_L|^2}\,,
\end{align}
in which the lattice momenta are
\begin{align}
 P_L&=\frac{1}{2i\sqrt{T_2U_2}}\left(m^2-\bar U m^1+\bar T(n_1+\bar U n_2)\right)\,,\\
 P_R&=\frac{1}{2i\sqrt{T_2U_2}}\left(m^2-\bar U m^1+T(n_1+\bar U n_2)\right)\,,
\end{align}
We have set the Wilson lines to zero for simplicity. The momenta satisfy
\begin{equation}
 |P_R|^2-|P_L|^2=n\cdot m=n_1m^1+n_2m^2\,.
\end{equation}
We now dress the lattice with some power of lattice momenta. For instance, consider including an even power of $Q_L$ in \eqref{LagLat}. This can be accounted for by deforming the latter with a factor of $e^{\lambda P_L}$ and then taking appropriate derivatives. Hence, we can use the same formula \eqref{poisson} to find
\begin{align}
 \frac{T_2}{\tau_2}\sum_{\tilde m_i,n_i}\left(\frac{Q_L}{\sqrt{\tau_2U_2}}\right)^{2k}q_T^{|Q_L|^2}\bar q_{\bar T}^{|Q_R|^2}=\sum_{m^i,n_i}\left(\frac{P_L}{\sqrt{T_2U_2}}\right)^{2k}q^{|P_L|^2}\bar q^{|P_R|^2}\,.
\end{align}
There are similar results for the other lattice momenta. More precisely, we have the following dictionary.
\begin{align}\label{Dict}
 P_L&\longleftrightarrow \,\,\,\,Q_L\,,\\
 P_R&\longleftrightarrow -\bar Q_R\,,\\
 \bar P_L&\longleftrightarrow \,\,\,\,Q_R\,,\\
 \bar P_R&\longleftrightarrow -\bar Q_L\,.
\end{align}

Consider now including different lattice momenta at the same time. For the case of interest, we consider an arbitrary power of $P_L\bar P_R$ and apply \eqref{poisson}. Naively, one would apply the dictionary \eqref{Dict} but there are subtleties due to the torus propagator $\left<X\bar X\right>$. Our starting point is the correlation function
\begin{equation}
 \left<(\partial X)^{k_1}(\bar\partial\bar X)^{k_2}\right>
\end{equation}
for arbitrary positive integers $k_1$ and $k_2$. At first sight it might appear that only zero modes contribute to the above correlator. However, on the world sheet torus (as well as on higher genera surfaces) there is a non-trivial correlation function due to the presence of zero modes of 1-differentials. Indeed, the torus  Green's function  for the Laplacian acting on scalars contains a term that goes as $\frac{1}{4\pi}\log |E(z,w)|^2$ where $E(z,w) = \theta_1(z-w)/\theta_1'(0)$. However this is not monodromy invariant under $z \rightarrow z+\tau$. The combination that is monodromy invariant is 
\begin{equation}
\frac{1}{4\pi}\log|E(z,w)|^2 - \frac{{\rm  Im}(z-w)^2}{2\tau_2}
\label{scalargreensfn}
\end{equation}
The correlator $\left<X(z)\bar{X}(w)\right>$ is therefore proportional to the above expression. As a result of the second  term in (\ref{scalargreensfn}), $\left<\partial X(z)\bar{\partial}\bar{X}(w)\right>$ is not zero, instead it is a constant. 

To proceed further, we define the generating function
\begin{equation}
 F_{\ell}(\lambda,\bar\lambda)=\sum_{k_1,k_2\in\mb N} \frac{\lambda^{k_1}}{k_1!}\frac{\bar\lambda^{k_2}}{k_2!}\left<(\partial X)^{k_1}(\bar\partial\bar X)^{k_2}\right>\,.
\end{equation}
Consider the case where there are $s$ operators $\partial X$ and $\bar\partial\bar X$ contracting. Then the remaining operators can only give zero-modes. Including the combinatorial factors of this choice of partition, and summing over $s$ yields
\begin{align}
 F_{\ell}(\lambda,\bar\lambda)=\sum_{k_1,k_2\in\mb N} \frac{\lambda^{k_1}}{k_1!}\frac{\bar\lambda^{k_2}}{k_2!}\sum_{s=0}^{\textrm{min}(k_1,k_2)}{k_1\choose s}{k_2\choose s}s!\left(\sqrt{\frac{T_2}{\tau_2}}Q_L\right)^{k_1-s}\left(\sqrt{\frac{T_2}{\tau_2}}\bar Q_L\right)^{k_2-s}\left(-\frac{1}{4\pi\tau_2}\right)^s\,.
\end{align}
Now defining $p=k_1-s$, $q=k_2-s$, we find
\begin{align}
 F_{\ell}(\lambda,\bar\lambda)&=\sum_{p,q,s\in\mb N} \frac{\lambda^{p+s}\bar\lambda^{q+s}}{p!q!s!}\left(\sqrt{\frac{T_2}{\tau_2}}Q_L\right)^{p}\left(\sqrt{\frac{T_2}{\tau_2}}\bar Q_L\right)^{q}\left(-\frac{1}{4\pi\tau_2}\right)^s\nonumber\\
                              &=\exp\left[\lambda\sqrt{\frac{T_2}{\tau_2}}Q_L+\bar\lambda\sqrt{\frac{T_2}{\tau_2}}\bar Q_L-\frac{\lambda\bar\lambda}{4\pi \tau_2}\right]\,.
\end{align}
We now perform the Poisson re-summation as above and find
\begin{align}
 \frac{T_2}{\tau_2}\sum_{\tilde m_i,n_i}e^{-2i\pi\bar T(\tilde m_1n_2-\tilde m_2n_1)-4\pi T_2|Q_L|^2+\lambda\sqrt{\frac{T_2}{\tau_2}}Q_L+\bar\lambda\sqrt{\frac{T_2}{\tau_2}}\bar Q_L-\frac{\lambda\bar\lambda}{4\pi\tau_2}}=\sum_{m^i,n_i}\bar q^{n\cdot m}e^{-4\pi\tau_2|P_L|^2+\lambda P_L-\bar\lambda\bar P_R}\,.
\end{align}
Expanding in the exponential back yields
\begin{align}
 \frac{T_2}{\tau_2}\sum_{\tilde m_i,n_i}\left<(\partial X)^{k_1}(\bar\partial\bar X)^{k_2}\right> q_T^{|Q_L|^2}\bar q_{\bar T}^{|Q_R|^2}=\sum_{m^i,n_i}(P_L)^{k_1}(-\bar P_R)^{k_2} q^{|P_L|^2}\bar q^{|P_R|^2}\,.
\end{align}
This is the result needed in order to `Gaussianise' the generating function \eqref{GenSim}.

\section{Functional determinants}\label{appendix:reg}

In this section, we show how to calculate the functional determinants arising from \eqref{GenSim} in a modular invariant fashion. The techniques involved are the same as the ones used in \cite{AFHNZ}, see also \cite{These}. Namely, we expand the operator in the exponential of \eqref{GenSim} in the eigenmodes of the Laplacian on the torus and then perform the integral over the fields $Z^\mu$. This leads to an infinite product that we define using $\zeta$-function regularisation after taking its logarithm. Explicitly, if we denote the bosonic part of \eqref{GenSim} by $F_{\textrm{bos}}$, then
\begin{align}
 \log(F_{\textrm{bos}})&=-\log\prod_{(m,n)\neq(0,0)}\left(|m+n\tau|^2+(m+n\tau)\tilde\epsilon+(m+n\bar\tau)\hat{\bar\epsilon}\right)^2\nonumber\\
&=2\lim_{s\rightarrow0}\frac{\partial}{\partial s}\sum\limits_{(m,n)\neq(0,0)}\sum_{k\geq0}\binom{-s}{k}\frac{1}{|m+n\tau|^{2s}}\left(\frac{\tilde\epsilon}{m+n\bar\tau}+\frac{\hat{\bar\epsilon}}{m+n\tau}\right)^k\nonumber\\
&=2\lim_{s\rightarrow0}\frac{\partial}{\partial s}\sum_{\genfrac{}{}{0pt}{}{k\geq0}{k\in2\mb Z}}\sum_{\genfrac{}{}{0pt}{}{0\leq \ell\leq k}{l\in2\mb Z}}\binom{-s}{k}\binom{k}{\ell}{\tilde\epsilon}^{\,\ell}\,{\hat{\bar\epsilon}}^{\,k-\ell}\,\tau_2^{\,\ell-k-s}\,\Phi^\ast_{k-l+s,k+2s}\,,\label{prescription}
\end{align}
where $\Phi_{\alpha,\beta}$ is the modular series of weight $(0,\beta-2\alpha)$ related to the usual Eisenstein series
\begin{equation}
 E(s,w)=\frac{1}{2}\sum_{(c,d)=1}\frac{\tau_2^{s-w/2}}{|c\tau+d|^{2s-w}}(c\tau+d)^{-w}
\end{equation}
through
\begin{equation}
 \Phi_{\alpha,\beta}=2\zeta(\beta)E(\beta/2,\beta-2\alpha)\,.
\end{equation}
It consistently defines the functional determinant for large $s$ and is absolutely convergent for $\beta>2$. The Fourier expansion of $\phi$ was derived in \cite{AFHNZ} and allows us to extract the asymptotic behaviour of $F_{\textrm{bos}}$. That is,
\begin{align}
 \lim_{\tau_2\rightarrow\infty}\log[F_{\textrm{bos}}]&=\zeta(2)(\tilde\epsilon^{\,2}+\hat{\bar\epsilon}^{\,2})+2\sum_{k\geq2}\frac{\zeta(2k)}{k}(\tilde\epsilon+\hat{\bar\epsilon})^{2k}\nonumber\\
 &=2\sum_{k\geq1}\frac{\zeta(2k)}{k}(\tilde\epsilon+\hat{\bar\epsilon})^{2k}\,.
\end{align}
Using the definition of the Riemann zeta function, the sum over $k$ can be performed and leads to the expected field theory limit:
\begin{align}
 \lim_{\tau_2\rightarrow\infty}\log[G^{\textrm{bos}}(\epsilon_-,\epsilon_+)]&=2\sum_{n,k\geq1}\frac{1}{k}\left(\frac{\tilde\epsilon+\hat{\bar\epsilon}}{n}\right)^{2k}=-2\sum_{n\geq1}\log\left[1-\left(\frac{\tilde\epsilon+\hat{\bar\epsilon}}{n}\right)^{2}\right]=-\log\left[\textrm{sinc}\,\pi(\tilde\epsilon+\hat{\bar\epsilon})^2\right]\,.
\end{align}
Finally, the path integral over the fermionic degrees of freedom in \eqref{GenSim} can be similarly derived and simply gives $\theta_1(\hat{\bar\epsilon})$ which reduces $\textrm{sinc}(\pi\hat{\bar\epsilon})^2$ in the field theory limit, thus obtaining the result in \eqref{wrongres}.

\bibliographystyle{utphys}
\bibliography{referencesNew}
\end{document}